\documentclass[11pt,a4paper]{article}
\usepackage[T1]{fontenc}
\usepackage[utf8]{inputenc}
\usepackage{amsmath,amssymb,mathtools}
\usepackage[margin=1in]{geometry}
\usepackage[expansion=false]{microtype}
\usepackage{hyperref}
\hypersetup{colorlinks=true,linkcolor=blue,citecolor=blue,urlcolor=blue}
\newcommand{\PV}{\operatorname{PV}}
\newcommand{\sgn}{\operatorname{sgn}}
\newcommand{\dd}{\mathrm{d}}
\newcommand{\F}{\mathcal F}
\newcommand{\Gfun}[2]{G_{0,3}^{2,0}\!\left(#1\,\middle|\begin{matrix}-\\#2\end{matrix}\right)}
\usepackage{lmodern}
\usepackage{authblk}

\title{Fourier Transform of the Scalar Propagator\\
	with Quantum Light-Cone Fluctuations}
\author[1,2]{M. V. Takook\thanks{\texttt{takook@apc.in2p3.fr}}}

\affil[1]{APC, UMR 7164, Universit\'e Paris Cit\'e, 75205 Paris, France}
\affil[2]{Quantum Statetronic, Bagneux, Hauts-de-Seine, France}
\date{\today}

\begin{document}
\maketitle

\begin{abstract}
	We derive the Lorentzian Fourier transform of the Gaussian-smeared light-cone contribution to the scalar propagator in Krein-space quantization in terms of a Meijer $G$-function. We characterize its distributional structure, recover the massive propagator by adding the Bessel-function tail, and derive an exact expression for the massless one-loop bubble kernel using a position-space prescription.
\end{abstract}

\vspace{0.5cm}


\section{Introduction}

Krein-space quantization introduces an indefinite inner product and auxiliary negative-norm degrees of freedom. For the scalar two-point function considered here, the resulting time-ordered kernel is the real, principal-value part of the conventional Feynman propagator \cite{ta02}. Metric fluctuations can broaden the classical light cone, motivating the replacement of its delta-function contribution by a Gaussian \cite{Ford,Takook}. We adopt a Gaussian of finite width, normalized as a distribution in the invariant interval.

The Fourier transform of this Gaussian is not an absolutely convergent four-dimensional integral, since the Gaussian remains constant along the noncompact surfaces of fixed Minkowski interval. Consequently, treating it as a Euclidean radial transform would alter the problem. We therefore compute the Lorentzian Fourier transform in four-dimensional Minkowski space in the sense of tempered distributions. An auxiliary Gaussian integral reduces the calculation to an Abel-summed sine integral, which we evaluate in terms of a Meijer $G$-function.

We obtain explicit expressions for timelike and spacelike momenta and determine the distributional extension at the null cone. This analysis separates the principal-value massless pole from a continuous remainder. Adding the Bessel-function tail yields the massive propagator and demonstrates the cancellation of the massless pole when the light-cone smearing has unit normalization. We then apply this result to derive an exact expression for the massless one-loop bubble kernel using a position-space prescription.

Throughout this work, we use a constant positive smearing parameter $\mu$, which characterizes the Gaussian width and is related to the metric-fluctuation variance. This prescription defines a model for the light-cone contribution rather than a general metric-averaged propagator. In a more general treatment, the variance may depend on the two spacetime points and on the gravitational state; here, we approximate it by a constant to obtain an analytically tractable model.


\section{Conventions and the smearing model}

We work in four-dimensional Minkowski spacetime with
\begin{equation}
	\eta_{\alpha\beta}
	=\operatorname{diag}(1,-1,-1,-1).
\end{equation}
We set
\begin{equation}
	y=x-x',\qquad q=p^2,\qquad
	y^2=(y^0)^2-|\mathbf y|^2,\qquad
	\sigma_0=\frac{y^2}{2}.
\end{equation}
Our Fourier-transform conventions are
\begin{equation}\label{eq:fourier}
	\widetilde f(p)
	=\int\mathrm{d}^4y\,e^{ip\cdot y}f(y),
	\qquad
	f(y)
	=\int\frac{\mathrm{d}^4p}{(2\pi)^4}
	e^{-ip\cdot y}\widetilde f(p),
\end{equation}
where $p\cdot y=p^0y^0-\mathbf p\cdot\mathbf y$.

The unsmeared scalar time-ordered kernel in
Krein-space quantization is defined by
\cite{ta02,Takook}
\begin{equation}
	G_{T,m}^{(0)}(y)
	=-i\langle0|T\phi(x)\phi(x')|0\rangle,
	\qquad
	\widetilde G_{T,m}^{(0)}(p)
	=\operatorname{PV}\frac{1}{p^2-m^2}.
\end{equation}
Its coordinate-space representation is
\begin{align}
	G_{T,m}^{(0)}(y)
	&=-\frac{1}{8\pi}\delta(\sigma_0)+B_m(y)
	=-\frac{1}{4\pi}\delta(y^2)+B_m(y),
	\label{eq:unsmeared}\\
	B_m(y)
	&=\frac{m^2}{8\pi}\theta(\sigma_0)
	\frac{J_1\!\left(\sqrt{2m^2\sigma_0}\right)}
	{\sqrt{2m^2\sigma_0}},
	\qquad m>0.
	\label{eq:tail}
\end{align}
The Bessel contribution is evaluated for
$\sigma_0>0$ and extended by zero for $\sigma_0<0$.
At the light cone, the ratio is defined by its
continuous limit $J_1(z)/z\to1/2$.
The value assigned to $\theta(0)$ does not affect
the distribution. For the massless field, we set
$B_0=0$.

Quantum metric fluctuations motivate a Gaussian
smearing of the light-cone contribution
\cite{Ford,Takook}. We introduce the normalized
Gaussian
\begin{equation}
	\delta_\mu(u)
	=\frac{1}{2\sqrt{\pi}\mu}
	\exp\!\left(-\frac{u^2}{4\mu^2}\right),
	\qquad
	\int_{\mathbb R}\delta_\mu(u)\,\mathrm{d}u=1,
	\qquad \mu>0.
\end{equation}
We smear the delta distribution in the Minkowski
interval according to
\begin{equation}
	\delta(y^2)\longrightarrow\delta_\mu(y^2).
\end{equation}
Since $\sigma_0=y^2/2$ and
$\delta_{\mu/2}(\sigma_0)=2\delta_\mu(y^2)$,
this prescription is equivalent to replacing
$\delta(\sigma_0)$ by
$\delta_{\mu/2}(\sigma_0)$ in
\eqref{eq:unsmeared}. The resulting propagator
model is
\begin{equation}\label{eq:model}
	\begin{aligned}
		G_{\mu,m}(y)&:=K_\mu(y)+B_m(y),\\
		K_\mu(y)&=-\frac{1}{8\pi\sqrt{\pi}\mu}
		\exp\!\left[-\frac{(y^2)^2}{4\mu^2}\right].
	\end{aligned}
\end{equation}
Here $(y^2)^2$ denotes the square of the Minkowski
interval, not a Euclidean radial norm. The
prescription smears only the light-cone contribution
and leaves the Bessel-function tail unchanged.

For $m>0$, the Fourier representation of the
Bessel contribution is
\begin{equation}\label{eq:tail-fourier}
	B_m(y)
	=
	\int\frac{\mathrm{d}^4p}{(2\pi)^4}
	e^{-ip\cdot y}
	\operatorname{PV}
	\frac{m^2}{p^2(p^2-m^2)}.
\end{equation}
The principal-part prescription in this expression
is defined by
\begin{equation}\label{eq:tail-pv}
	\operatorname{PV}
	\frac{m^2}{p^2(p^2-m^2)}
	:=
	\operatorname{PV}\frac{1}{p^2-m^2}
	-\operatorname{PV}\frac{1}{p^2}.
\end{equation}
This specifies the distribution through its
partial-fraction decomposition.

The Fourier transform of $K_\mu$ is defined by
\begin{equation}\label{eq:first-fourier}
	K_\mu(y)
	=\int\frac{\mathrm{d}^4p}{(2\pi)^4}
	e^{-ip\cdot y}\widetilde K_\mu(p).
\end{equation}
Consequently,
\begin{equation}\label{eq:full-fourier}
	\begin{aligned}
		G_{\mu,m}(y)
		&=\int\frac{\mathrm{d}^4p}{(2\pi)^4}
		e^{-ip\cdot y}\widetilde G_{\mu,m}(p),\\
		\widetilde G_{\mu,m}(p)
		&=\widetilde K_\mu(p)
		+\operatorname{PV}
		\frac{m^2}{p^2(p^2-m^2)},
		\qquad m>0.
	\end{aligned}
\end{equation}
For $m=0$, the Bessel contribution vanishes and
$\widetilde G_{\mu,0}=\widetilde K_\mu$.
All Fourier transforms are understood in the
sense of tempered distributions.

In units $\hbar=c=1$, the dimensions are
\begin{equation}
	[\mu]=\mathrm{mass}^{-2},\qquad
	[G_{\mu,m}]=\mathrm{mass}^{2},\qquad
	[\widetilde G_{\mu,m}]=\mathrm{mass}^{-2}.
\end{equation}
If the first-order interval fluctuation $\sigma_1$
is modeled as a centered Gaussian random variable
with constant variance
$v=\langle\sigma_1^2\rangle>0$, then
\begin{equation}
	\left\langle\delta(\sigma_0+\sigma_1)\right\rangle
	=\frac{1}{\sqrt{2\pi v}}
	\exp\!\left(-\frac{\sigma_0^2}{2v}\right).
\end{equation}
Under this Gaussian model, averaging broadens
the light-cone contribution
\cite{Ford}. Comparison with the smearing in
\eqref{eq:model} gives $\mu^2=2v$. The variance
depends on the gravitational state and its
correlations; a quantitative relation to graviton
density requires specifying that state. Here and
below, single angle brackets
$\langle\cdot\rangle$ denote expectation values
over metric fluctuations.

The model \eqref{eq:model} is an approximation to
the metric-averaged propagator, rather than a
general expression for that average. The normalized
smearing ensures the distributional zero-width limit
\begin{equation}\label{eq:zerowidth}
	G_{\mu,m}\longrightarrow G_{T,m}^{(0)}
	\quad\text{in }\mathcal S'(\mathbb R^4),
	\qquad \mu\downarrow0.
\end{equation}
Here $\mathcal S'(\mathbb R^4)$ denotes the space
of tempered distributions, the continuous dual
of the Schwartz space $\mathcal S(\mathbb R^4)$
of smooth functions whose derivatives of every
order decrease faster than any inverse power
at infinity. The convergence means that
\begin{equation}
	\lim_{\mu\downarrow0}G_{\mu,m}[\varphi]
	=G_{T,m}^{(0)}[\varphi]
	\qquad
	\text{for every }\varphi\in\mathcal S(\mathbb R^4),
\end{equation}
where $G[\varphi]$ denotes the action of the
distribution $G$ on the test function $\varphi$.
For the bounded function $G_{\mu,m}$, this action
is given by
\begin{equation}
	G_{\mu,m}[\varphi]
	=\int_{\mathbb R^4}\mathrm{d}^4y\,
	G_{\mu,m}(y)\varphi(y).
\end{equation}

The alternative Gaussian coefficient
$-\frac{1}{8\pi}\sqrt{\frac{\pi}{2\mu^2}}$
multiplies $K_\mu$ by $\pi/\sqrt{2}$.
It therefore gives a different zero-width
normalization when the Bessel tail is unchanged.
We use \eqref{eq:model} consistently below;
Appendix~\ref{app:normalization} records the
conversion to the alternative coefficient.

\section{Distributional Lorentzian Fourier transform}

Since $K_\mu$ is bounded, it defines a tempered
distribution on Minkowski spacetime
$\mathbb R^{1,3}$. Its Fourier transform can be
obtained by Gaussian regularization:
\begin{equation}\label{eq:abel4}
	\widetilde K_\mu(p)
	=
	\lim_{\varepsilon\downarrow0}
	\int_{\mathbb R^{1,3}}\mathrm{d}^4y\,
	e^{ip\cdot y}
	e^{-\varepsilon[(y^0)^2+|\mathbf y|^2]}
	K_\mu(y),
\end{equation}
where the limit is taken in
$\mathcal S'(\mathbb R^4)$.
The integration variables remain real.
The positive-definite Gaussian regulator ensures
absolute convergence for $\varepsilon>0$ without
a Wick rotation. Although the regulator breaks
Lorentz invariance at finite $\varepsilon$,
the limiting Fourier transform is Lorentz invariant
because $K_\mu$ depends only on $y^2$.

We use the auxiliary identity
\begin{equation}
	e^{-(y^2)^2/(4\mu^2)}
	=\frac{1}{\sqrt{\pi}}
	\int_{-\infty}^{\infty}\mathrm{d}s\,
	e^{-s^2}e^{is y^2/\mu}.
\end{equation}
With the regulator retained, the coordinate
integrations are Gaussian. For
$\alpha\in\mathbb R\setminus\{0\}$, the
corresponding distributional Fresnel limit is
\begin{equation}\label{eq:fresnel}
	\int\mathrm{d}^4y\,
	e^{ip\cdot y+i\alpha y^2}
	=-i\pi^2\frac{\operatorname{sgn}(\alpha)}
	{\alpha^2}
	e^{-ip^2/(4\alpha)}.
\end{equation}
The phase factor follows from one time integration
and three spatial integrations. Evaluating the
coordinate integrals at finite $\varepsilon$ and
then taking the distributional limit gives
\begin{equation}
	\widetilde K_\mu(p)
	=\frac{i\mu}{8}
	\int_{-\infty}^{\infty}\mathrm{d}s\,
	\frac{\operatorname{sgn}(s)}{s^2}
	e^{-s^2-i\mu q/(4s)}.
\end{equation}
The integral at $s=0$ inherits its prescription
from the original regulated Fourier transform;
it is not an ordinary improper integral.

Pairing the contributions from positive and
negative $s$, and then setting $t=1/s$ on the
positive half-line, yields
\begin{equation}\label{eq:sine}
	\boxed{\widetilde K_\mu(p)=\frac{\mu}{4}I\!\left(\frac{\mu q}{4}\right),\;\; \;\;
I(a)=\operatorname{Abel}\int_0^\infty\mathrm{d}t\,e^{-1/t^2}\sin(at).	}
\end{equation}
Here Abel summation means inserting $e^{-\eta t}$
and taking $\eta\downarrow0$ in the sense of
tempered distributions. The unregulated sine
integral is not an ordinary improper integral
because its amplitude tends to one at infinity.

A useful distributional decomposition is
\begin{equation}\label{eq:subtract}
	I(a)
	=\operatorname{PV}\frac{1}{a}
	+\int_0^\infty\mathrm{d}t\,
	\bigl(e^{-1/t^2}-1\bigr)\sin(at).
\end{equation}
The first term follows from
\begin{equation}
	\int_0^\infty e^{-\eta t}\sin(at)\,\mathrm{d}t
	=\frac{a}{a^2+\eta^2}
	\longrightarrow
	\operatorname{PV}\frac{1}{a}
	\quad\text{in }\mathcal S'(\mathbb R),
	\qquad \eta\downarrow0.
\end{equation}
The amplitude in the second integral is integrable:
\begin{equation}
	\int_0^\infty
	\left|e^{-1/t^2}-1\right|\mathrm{d}t<\infty.
\end{equation}
Consequently, the second term in
\eqref{eq:subtract} defines a continuous odd
function of $a$.

To justify the corresponding decomposition in
four-momentum space, subtract the unsmeared
massless light-cone kernel before removing the
regulator in \eqref{eq:abel4}. In the auxiliary
representation, this subtraction replaces
$e^{-s^2}$ by $e^{-s^2}-1$.
After pairing with a Schwartz function, the regulated
coordinate integrals are bounded uniformly by a constant
times $|s|^{-2}$ for $s\ne0$: the modulus of the
regulated Gaussian prefactor is at most $\pi^2\mu^2/s^2$,
and the remaining exponential has modulus at most one.
Thus the subtracted integrand has an integrable
majorant proportional to
\begin{equation}
	\frac{|e^{-s^2}-1|}{s^2},
	\qquad
	\int_{\mathbb R}
	\frac{|e^{-s^2}-1|}{s^2}\,\mathrm{d}s<\infty.
\end{equation}
Dominated convergence therefore applies to the
subtracted integral after pairing with Schwartz
test functions. The unsmeared massless light-cone
term is fixed by \eqref{eq:unsmeared} and transforms
to $\operatorname{PV}(1/p^2)$. Thus \eqref{eq:sine} is understood explicitly as
\begin{equation}
	\widetilde K_\mu(p)
	=
	\operatorname{PV}\frac{1}{p^2}
	+\frac{\mu}{4}
	\int_0^\infty\mathrm{d}t\,
	\bigl(e^{-1/t^2}-1\bigr)
	\sin\!\left(\frac{\mu p^2t}{4}\right).
\end{equation}
Here
\begin{equation}
	\operatorname{PV}\frac{1}{p^2}
	:=
	\lim_{\rho\downarrow0}
	\frac{p^2}{(p^2)^2+\rho^2}
	\quad\text{in }\mathcal S'(\mathbb R^4).
\end{equation}
This separates the principal-value massless pole
from a continuous remainder, which vanishes at
$p^2=0$, and fixes the transform without additional
terms supported on the null cone. All exchanges
of integrations are understood with the stated
regulators or after pairing with Schwartz test
functions.


\section{Meijer \texorpdfstring{$G$}{G} representation}
For $a>0$, integration by parts in the Abel sense gives an absolutely convergent cosine integral:
\begin{equation}\label{eq:derivative}
	I(a)=\frac1a\int_0^\infty f'(t)\cos(at)\,\dd t,
	\qquad f(t)=e^{-1/t^2},\quad f'(t)=2t^{-3}e^{-1/t^2}.
\end{equation}
Define $\mathcal M[I](s)=\int_0^\infty a^{s-1}I(a)\,\dd a$. Using a regulator for the oscillatory interchange gives
\begin{align}
	\mathcal M[I](s)
	&=\Gamma(s-1)\cos\!\left(\frac{\pi(s-1)}2\right)
	\int_0^\infty f'(t)t^{1-s}\,\dd t\notag\\
	&=\frac12\Gamma(s)\Gamma\!\left(\frac{s-1}{2}\right)
	\sin\!\left(\frac{\pi s}{2}\right),\qquad 1<\Re s<2.\label{eq:mellin}
\end{align}
This route avoids using the elementary Mellin transform of the sine outside its initial strip of validity. With $s=2u$, the duplication and reflection identities give
\begin{equation}\label{eq:MB}
	I(a)=\frac{\sqrt\pi}{2}\frac{1}{2\pi i}
	\int_{c-i\infty}^{c+i\infty}
	\frac{\Gamma(u-\tfrac12)\Gamma(u+\tfrac12)}{\Gamma(1-u)}
	\left(\frac{a^2}{4}\right)^{-u}\dd u,\qquad \frac12<c<1.
\end{equation}
Comparison with the Mellin--Barnes definition \cite{DLMF} gives
\begin{equation}
	I(a)=\frac{\sqrt\pi}{2}\Gfun{a^2/4}{-\tfrac12,\tfrac12,0},\qquad a>0.
\end{equation}
Oddness supplies the result for negative $a$. Consequently, away from $q=0$,
\begin{equation}\label{eq:Kmeijer}
	\boxed{\quad\widetilde K_\mu(p)=\sgn(q)\frac{\sqrt\pi\mu}{8}
		\Gfun{\mu^2q^2/64}{-\tfrac12,\tfrac12,0}.\quad}
\end{equation}
The factor $\sgn(q)$ is essential. Dependence on $q^2$ alone would give the wrong spacelike sign.

Define the dimensionless function
\begin{equation}\label{eq:H}
	H(z)=\Gfun{z}{0,1,\tfrac12},\qquad z>0.
\end{equation}
The shift identity for the Meijer $G$-function gives the equivalent pointwise formula
\begin{equation}\label{eq:compact}
	\widetilde K_\mu(p)=\frac{\sqrt\pi}{q}H\!\left(\frac{\mu^2q^2}{64}\right),\qquad q\ne0.
\end{equation}
At the null cone, the Fourier transform determined by \eqref{eq:abel4} has the explicit decomposition
\begin{equation}\label{eq:R}
	\boxed{\quad\widetilde K_\mu(p)=\PV\frac1{p^2}+R_\mu(p^2),\qquad
		R_\mu(q)=\frac{\sqrt\pi H(\mu^2q^2/64)-1}{q},\quad R_\mu(0)=0.\quad}
\end{equation}
Here $\PV(p^2)^{-1}$ is the real part of the distributional boundary value of $(p^2+i0)^{-1}$. Equation \eqref{eq:R}, equivalently \eqref{eq:subtract}, specifies the meaning of the special-function expression at $p^2=0$.

Subtracting the massless version of \eqref{eq:unsmeared} from the massive one fixes the tail transform without an additional integration:
\begin{equation}\label{eq:Btransform}
	\widetilde B_m(p)=\PV\frac1{p^2-m^2}-\PV\frac1{p^2}.
\end{equation}
The frequently used notation $\PV\{m^2/[p^2(p^2-m^2)]\}$ means precisely this difference, not an unspecified product of singular distributions. Combining \eqref{eq:R} and \eqref{eq:Btransform}, we obtain
\begin{equation}\label{eq:main}
	\boxed{\quad\widetilde G_{\mu,m}(p)=\PV\frac1{p^2-m^2}+R_\mu(p^2),\qquad m\ge0.\quad}
\end{equation}
For $m>0$, the massless pole cancels exactly. For $m=0$, the massless pole remains. The massive physical pole at $p^2=m^2$ is also retained. A principal-value pole is a well-defined distribution, but it is not a bounded function. Both limiting operations are consistent:
\begin{equation}
	\lim_{\mu\downarrow0}\widetilde G_{\mu,m}=\PV\frac1{p^2-m^2},\qquad
	\lim_{m\downarrow0}\widetilde G_{\mu,m}=\widetilde K_\mu,
\end{equation}
in $\mathcal S'(\mathbb R^4)$. The second limit is not obtained by taking a fixed-$m$ infrared expansion and then setting $m=0$.


\section{Infrared and large-invariant-momentum behavior}
\subsection{Expansion near the null cone}
The coincident poles in the Mellin--Barnes representation of $H$ generate logarithms. Its residue expansion is
\begin{equation}\label{eq:series}
	H(z)=\frac1{\sqrt\pi}+\sum_{n=1}^\infty
	\frac{z^n\{\ln z-\psi(n+1)-\psi(n)-\psi(n+\tfrac12)\}}
	{n!(n-1)!\Gamma(n+\tfrac12)},\qquad z>0,
\end{equation}
where $\psi$ is the digamma function. In particular, as $z\downarrow0$,
\begin{equation}
	\sqrt\pi H(z)=1+2z\,[\ln z-3+3\gamma+2\ln2]
	+O(z^2|\ln z|).
\end{equation}
Here $\gamma$ denotes the Euler--Mascheroni constant. Thus, for fixed $\mu$ and $q\to0$,
\begin{equation}\label{eq:Rsmall}
	R_\mu(q)=\frac{\mu^2q}{32}
	\left[\ln\!\left(\frac{\mu^2q^2}{64}\right)-3+3\gamma+2\ln2\right]
	+O\!\left(\mu^4q^3|\ln(\mu^2q^2)|\right).
\end{equation}
This proves that $R_\mu$ is continuous at $q=0$ but is not differentiable there. For $m>0$ and $|q|\ll m^2$,
\begin{equation}
	\widetilde G_{\mu,m}(p)=-\frac1{m^2}-\frac{q}{m^4}
	+O(q^2/m^6)+R_\mu(q).
\end{equation}
For $m=0$, the leading term is instead $\PV(1/q)$. Accordingly, smearing does not remove the massless propagator pole. Whether a particular amplitude has an infrared divergence depends on its full integrand and external kinematics.

\subsection{Large invariant momentum}
The function $f'(t)$ in \eqref{eq:derivative} is flat at $t=0$, and all its derivatives are integrable and vanish at both integration endpoints. Repeated integration by parts therefore shows that, for every integer $N>0$,
\begin{equation}
	I(a)=O(|a|^{-N}),\qquad |a|\to\infty.
\end{equation}
Consequently $\widetilde K_\mu$ decreases faster than every inverse power of $|q|$ for fixed $\mu>0$. No particular exponential rate is needed for this conclusion. The massive tail then determines the algebraic asymptotics:
\begin{equation}\label{eq:UV}
	\widetilde G_{\mu,m}(p)=\frac{m^2}{q^2}+\frac{m^4}{q^3}
	+O(q^{-4}),\qquad |q|\to\infty,\quad m>0.
\end{equation}
This is an expansion away from the pole surfaces. Large $|p^2|$ is not the same as large Euclidean norm of the four-momentum: arbitrarily large momenta can remain close to the light cone.


\section{Coordinate-space regularity and a massless bubble kernel}

At fixed $\mu>0$, $K_\mu$ is smooth and bounded.
For $m>0$, the Bessel-function tail $B_m(y)$ is
bounded but has a finite jump across the light cone.
As $y^2\downarrow0$,
\begin{equation}
	B_m(y)
	=\frac{m^2}{16\pi}
	-\frac{m^4y^2}{128\pi}
	+O\!\left(m^6(y^2)^2\right),
	\qquad y^2>0,
\end{equation}
whereas $B_m(y)=0$ for $y^2<0$.
For fixed $m>0$ and $\sigma_0\to+\infty$,
\begin{equation}
	B_m(y)
	=
	\frac{2^{-1/4}m^{1/2}}{8\pi\sqrt{\pi}}
	\sigma_0^{-3/4}
	\left[
	\cos\!\left(m\sqrt{2\sigma_0}-\frac{3\pi}{4}\right)
	+O\!\left((m\sqrt{2\sigma_0})^{-1}\right)
	\right].
\end{equation}
Thus the full kernel has finite one-sided limits at
the light cone, although it is not continuous there
for $m>0$. Its value on the light cone, including
at coincidence, depends on the convention chosen
for the jump. This choice does not affect the
associated distribution.

For a concrete application, let $P$ denote the
external four-momentum flowing through the bubble,
and set $Q=P^2$. We define the massless bubble kernel
as the Fourier transform of an ordinary bounded
product:
\begin{equation}\label{eq:bubbledef}
	\mathcal B_\mu(P):=\F[K_\mu^2](P),
\end{equation}
where the Fourier transform is understood in
$\mathcal S'(\mathbb R^4)$. Direct multiplication gives
\begin{equation}
	K_\mu(y)^2
	=\frac{1}{64\pi^3\mu^2}
	e^{-(y^2)^2/(2\mu^2)}
	=-\frac{1}{8\sqrt{2}\,\pi^{3/2}\mu}
	K_{\mu/\sqrt{2}}(y).
\end{equation}
Hence the exact result in this prescription is
\begin{equation}\label{eq:bubbleresult}
	\boxed{
		\mathcal B_\mu(P)
		=-\frac{1}{8\sqrt{2}\,\pi^{3/2}\mu}
		\widetilde K_{\mu/\sqrt{2}}(P).
	}
\end{equation}
For $Q\ne0$, this becomes
\begin{equation}
	\mathcal B_\mu(P)
	=-\frac{\sgn(Q)}{128\pi}
	\Gfun{\mu^2Q^2/128}{-\tfrac12,\tfrac12,0}.
\end{equation}
The bubble kernel is dimensionless, as required.
Its distributional extension at the null cone is
\begin{equation}
	\mathcal B_\mu(P)
	=-\frac{1}{8\sqrt{2}\,\pi^{3/2}\mu}
	\left[
	\PV\frac{1}{P^2}
	+R_{\mu/\sqrt{2}}(P^2)
	\right].
\end{equation}
In particular, it retains a principal-value
singularity at $P^2=0$.

One may write the symbolic momentum-space convolution
\begin{equation}
	\mathcal B_\mu(P)
	\equiv
	\int\frac{\dd^4k}{(2\pi)^4}
	\widetilde K_\mu(k)\widetilde K_\mu(P-k),
\end{equation}
provided the right-hand side is understood through
\eqref{eq:bubbledef}, or a compatible limiting
prescription. Here $k$ denotes the loop momentum,
so the two internal lines carry momenta $k$ and
$P-k$. A convolution of arbitrary tempered
distributions is not automatically defined by an
ordinary integral.

For the interaction
$\mathcal L_{\mathrm{int}}=-\lambda\phi^4/4!$,
we use vertex factor $-i\lambda$ and internal-line
factor $i\widetilde K_\mu$. We adopt the convention
$S=1+iT$ and define the invariant scattering
amplitude $\mathcal M_{fi}$ by
\begin{equation}
	\langle f|iT|i\rangle
	=i(2\pi)^4\delta^{(4)}(P_f-P_i)\mathcal M_{fi},
\end{equation}
where $P_i$ and $P_f$ are the total initial and final
four-momenta. With these conventions, the formal
amputated $s$-channel contribution is
\begin{equation}
	i\mathcal M_s^{(1)}(P)
	=\frac{(-i\lambda)^2i^2}{2}\mathcal B_\mu(P)
	=\frac{\lambda^2}{2}\mathcal B_\mu(P),
\end{equation}
where $1/2$ is the diagram's symmetry factor.
Interpreting this expression as a physical
scattering amplitude additionally requires
specifying the interacting Krein-space rules and
the physical-state prescription.

For the massive case, the same position-space
prescription gives
\begin{equation}
	\F[G_{\mu,m}^2](P)
	=
	\mathcal B_\mu(P)
	+2\F[K_\mu B_m](P)
	+\F[B_m^2](P).
\end{equation}
All three coordinate-space products are bounded
and define tempered distributions. Only the first
term has been reduced here to the displayed
Meijer $G$-function. Moreover, multiplying averaged
propagators is a specified effective prescription:
in general,
\begin{equation}
	\langle G(y)^2\rangle
	\ne
	\langle G(y)\rangle^2
\end{equation}
when the same fluctuating metric enters both
factors.


\section{Conclusion}

The Gaussian light-cone model admits an exact Lorentzian Fourier transform. Its Meijer $G$-function representation must include the factor $\operatorname{sgn}(p^2)$ (the timelike--spacelike sign) and a specified extension at the null cone. With unit-normalized light-cone smearing, the massive result is the principal-value massive pole plus a continuous correction; the massless result retains its principal-value massless pole. For fixed $\mu>0$, the Gaussian contribution decays faster than any inverse power of $|p^2|$ as $|p^2|\to\infty$, while the massive tail has the algebraic behavior $m^2/(p^2)^2$.

At fixed positive smearing width, the bounded coordinate-space kernels allow ordinary products for the nonderivative examples considered here. The exact bubble illustrates both the usefulness and the limits of that statement: its Fourier transform is well defined as a tempered distribution, although it retains a principal-value singularity at $P^2=0$. To establish ultraviolet and infrared finiteness and determine whether renormalization conditions are required, further work should specify the metric correlations, the averaging of products, and the interacting physical-state prescription before extending the calculation to general amplitudes.

\section*{Acknowledgements}

The author acknowledges ChatGPT for its assistance in discussions of the Meijer \(G\)-function.


\appendix
\section{Alternative Gaussian normalizations}\label{app:normalization}

For $\mu>0$, we consider the alternative Gaussian term
\begin{equation}
	K^{\mathrm{alt}}_\mu(y)=-\frac{1}{8\pi}\sqrt{\frac{\pi}{2\mu^2}}
	e^{-(y^2)^2/(4\mu^2)}=C K_\mu(y),\qquad C=\frac{\pi}{\sqrt2}.
\end{equation}
Keeping this coefficient is mathematically possible, but its zero-width limit is $-C\delta(y^2)/(4\pi)$, rather than the light-cone term in \eqref{eq:unsmeared}. Its exact transform is
\begin{equation}
	\widetilde K^{\mathrm{alt}}_\mu(p)=\sgn(q)\frac{\pi^{3/2}\mu}{8\sqrt2}
	\Gfun{\mu^2q^2/64}{-\tfrac12,\tfrac12,0},\qquad q\ne0.
\end{equation}
If the Bessel tail is left unchanged, the full transform
with the alternative normalization is
\begin{equation}
	\widetilde G^{\mathrm{alt}}_{\mu,m}(p)
	=\PV\frac1{p^2-m^2}
	+(C-1)\PV\frac1{p^2}
	+C R_\mu(p^2).
\end{equation}
For $m>0$, the massless pole therefore cancels only
when $C=1$. Equation \eqref{eq:model} adopts unit
normalization to recover the stated unsmeared
propagator. This changes the model normalization;
it is not an algebraic rearrangement of the
alternative coefficient.

The coefficient in Eq.~(2.5) of Ref.~\cite{Takook} must be compared using the variance appearing there. Writing $v=\langle\sigma_1^2\rangle$ and $\mu^2=2v$, its light-cone term is
\begin{equation}
	-\frac{1}{8\pi}\sqrt{\frac{\pi}{2v}}
	e^{-\sigma_0^2/(2v)}=\pi K_\mu(y).
\end{equation}
Thus that printed coefficient differs by a factor $\pi$ from the unit-normalized Gaussian average adopted here. The factor $C=\pi/\sqrt2$ above instead concerns the alternative expression written directly in terms of $\mu$ with exponent $-(y^2)^2/(4\mu^2)$. For the coefficient printed in Ref.~\cite{Takook}, the preceding conversion formulas apply with $C=\pi$ after matching the widths.


\section{Notation and computational checks}

The Meijer $G$-function $H(z)$ is represented in grouped
computational notation as
\begin{equation}
	H(z)=\Gfun{z}{0,1,\tfrac12}
	\equiv\operatorname{MeijerG}\!\left(
	((),()),\,((0,1),(\tfrac12)),\,z
	\right),
	\qquad z>0.
\end{equation}
There are no upper parameters. The lower parameters are
divided into two groups: $0$ and $1$ contribute the
numerator factors $\Gamma(u)\Gamma(u+1)$, whereas
$\tfrac12$ contributes the denominator factor
$\Gamma(\tfrac12-u)$. Explicitly,
\begin{equation}
	H(z)=\frac{1}{2\pi i}
	\int_{c_H-i\infty}^{c_H+i\infty}
	\frac{\Gamma(u)\Gamma(u+1)}
	{\Gamma(\tfrac12-u)}
	z^{-u}\,\mathrm{d}u,
	\qquad c_H>0,\quad z>0.
\end{equation}
Here $c_H$ is a positive real constant specifying the
vertical contour $\operatorname{Re}u=c_H$ in the complex
$u$-plane. The contour lies to the right of all poles of
$\Gamma(u)$ and $\Gamma(u+1)$.

For a numerical check of \eqref{eq:Kmeijer}, one may
compare the Meijer $G$-function expression for $I(a)$
with the convergent cosine integral
\eqref{eq:derivative}, rather than truncating the
nonconvergent, unregularized sine integral. Near $q=0$,
the expansion \eqref{eq:Rsmall} can be used to avoid
loss of numerical precision caused by subtractive
cancellation.

\end{document}